\documentclass[12pt]{article}
\usepackage{amsmath}
\usepackage{epsfig}
\usepackage{amssymb}
\input{epsf}
\newcommand\as{\alpha_{\mathrm{S}}}
\newcommand\f[2]{\frac{#1}{#2}}
\def\la{\lambda} 
 
\def\beq{\begin{equation}}
\def\eeq{\end{equation}}
\def\beeq{\begin{eqnarray}}
\def\eeeq{\end{eqnarray}}
\def\to{\rightarrow}
\def\nn{\nonumber}

\def\b0{b_0}
\def\bone{b_1}

\def\GE{\gamma_E}

\begin{document}

\begin{titlepage}
\renewcommand{\thefootnote}{\fnsymbol{footnote}}
\begin{flushright}
BNL-NT-07/17 \\    
hep-ph/yymmnnn
     \end{flushright}
\par \vspace{10mm}
\begin{center}
{\Large \bf
Resummed Cross Section for Jet Production\\[5mm] at Hadron Colliders}

\end{center}
\par \vspace{2mm}

\begin{center}
{\bf Daniel de Florian}

\vspace{3mm}

Departamento de F\'\i sica, FCEYN, Universidad de Buenos Aires,\\
(1428) Pabell\'on 1 Ciudad Universitaria, Capital Federal, Argentina

\vspace{8mm}
{\bf Werner Vogelsang}

\vspace{3mm}

BNL Nuclear Theory,
Brookhaven National Laboratory, Upton, NY 11973, USA

\end{center}

\par \vspace{2mm}
\begin{center} {\large \bf Abstract} \end{center}
\begin{quote}
\pretolerance 10000
We study the resummation of large logarithmic perturbative 
corrections to the single-inclusive jet cross section at hadron 
colliders. The corrections we address arise near the threshold 
for the partonic reaction, when the incoming partons have just
enough energy to produce the high-transverse-momentum final state. 
The structure of the resulting logarithmic corrections is known to 
depend crucially on the treatment of the invariant mass of the 
produced jet at threshold. We allow the jet to have a non-vanishing
mass at threshold, which most closely corresponds to the situation 
in experiment. Matching our results to available semi-analytical 
next-to-leading-order calculations, we derive resummed results valid
to next-to-leading logarithmic accuracy. We present numerical results 
for the resummation effects at Tevatron and RHIC energies.

\end{quote}

\vspace*{\fill}
\begin{flushleft}
     hep-ph/yymmnnn \\ 
\today

\end{flushleft}
\end{titlepage}

\setcounter{footnote}{1}
\renewcommand{\thefootnote}{\fnsymbol{footnote}}

\section{Introduction}

High-transverse-momentum jet production in hadronic collisions,
$H_1 H_2 \to {\mathrm{jet}}\,X$, 
plays a fundamental role in High-Energy Physics. It offers 
possibilities to explore QCD, for example the structure of the 
interacting hadrons or the emergence of hadronic final states, 
but is also intimately involved in many signals (and their backgrounds) 
for New Physics. At the heart of all these applications of jet production 
is our ability to perform reliable and precise perturbative calculations 
of the partonic short-distance interactions that generate the 
high-transverse-momentum final states. Up to corrections suppressed 
by inverse powers of the jet's transverse momentum $p_T$, the hadronic 
jet cross section factorizes into parton distribution functions that 
contain primarily long-distance information, and these short-distance 
cross sections. In the present paper, we address large logarithmic 
perturbative corrections to the latter.

At partonic threshold, when the initial partons have 
just enough energy to produce the high-$p_T$ jet
and an unobserved recoiling partonic final state, the 
phase space available for gluon bremsstrahlung vanishes, so
that only soft and collinear emission is allowed, 
resulting in large logarithmic corrections to the partonic cross 
section. To be more specific, if we consider the cross section 
as a function of the jet transverse momentum, integrated over all 
jet rapidities, the partonic threshold is reached when 
$\sqrt{s}=2 p_T$, where $\sqrt{s}$ is the partonic 
center-of-mass (c.m.) energy. Defining $\hat{x}_T\equiv 2 p_T/\sqrt{s}$,
the leading large contributions near threshold arise as $\as^k(p_T)
\ln^{2m}\left(1-\hat{x}_T^2\right)$ at the $k$th order in 
perturbation theory, where $m\leq k$ (the logarithms with $m=k$ are leading)
and $\as$ is the strong coupling. 
Even if $p_T$ is large so that $\as(p_T)$ is small, sufficiently
close to threshold the logarithmic terms will spoil the perturbative
expansion to any fixed order. Threshold 
resummation ~\cite{dyresum,Catani:1996yz,KS,KOS,KOS1,LOS,BCMN},
however, allows to reinstate a useful perturbative series by 
systematically taking into account the terms $\as^k \ln^{2m}
\left(1-\hat{x}_T^2\right)$ to all orders in $\as$. This is achieved
after taking a Mellin-transform of the hadronic cross section in 
$x_T=2 p_T/\sqrt{S}$, with $\sqrt{S}$ the hadronic c.m. energy. The 
threshold logarithms exponentiate in transform space. 

Regarding phenomenology, the larger $x_T$, the more dominant will the 
threshold logarithms be, and hence the more important will threshold 
resummation effects be. In addition, because of the convoluted form 
of the partonic cross sections and the parton distribution functions 
(PDFs), the steep fall-off of the PDFs with momentum fraction $x$
automatically enhances the contributions from the threshold regime
to the cross section, because it makes it relatively unlikely that the
initial partons have very high c.m. energy. This explains why partonic
threshold effects often dominate the hadronic cross section even 
at not so high $x_T$. Studies of  cross sections for~\cite{deFlorian:2005yj} 
$pp\to hX$ (with $h$ a high-$p_T$ hadron) and~\cite{CMN,KOgamma,sv,DdFWV} 
$pp\to \gamma 
X$ in the fixed-target regime, where typically $0.2\lesssim x_T \lesssim 
0.7$, indeed demonstrate that threshold-resummation effects dominate there
and can be very large and important for phenomenology. They enhance 
the theoretical cross section with respect to fixed-order calculations. 

These observations suggest to study the resummation also for jet production 
at hadron colliders, in particular when $x_T$ is rather large. An application
of particular interest is the jet cross section at very high transverse
momenta ($p_T\sim$~several hundreds GeV) at the 
Tevatron~\cite{Affolder:2001fa,Abazov:2001hb}, for which 
initially an excess of the experimental data over next-to-leading
order (NLO) theory was reported, which was later mostly attributed to
an insufficient knowledge of the gluon distribution~\cite{cteqjets}. 
Similarly large values of $x_T$ are now probed in $pp$ collisions at RHIC, 
where currently $\sqrt{s}=200$~GeV and jet cross section measurements 
by the STAR collaboration are already extending to 
$p_T\gtrsim 40$~GeV~\cite{Abelev:2006uq}.
In both these cases, one does expect threshold resummation effects 
to be smaller than in the case of related processes at similar $x_T$ in 
the fixed-target regime, just because (among other things) the strong 
coupling constant is smaller at these higher $p_T$. On the other hand,
as we shall see, the effects are still often non-negligible.

Apart from addressing these interesting phenomenological applications, 
we believe we also improve in this paper the theoretical framework 
for threshold resummation in jet production. There has been earlier 
work in the literature on this topic~\cite{KOS,KO,Bauer:2006qp}. 
In Ref.~\cite{KOS} the threshold resummation formalism for the closely 
related dijet production at large invariant mass of the jet pair
was developed to next-to-leading logarithmic (NLL) order. In~\cite{KO}, 
these results were applied to the single-inclusive jet cross section at 
large transverse momentum, making use of earlier work~\cite{LOS} on the 
high-$p_T$ prompt-photon cross section, which is kinematically similar.
As was emphasized in~\cite{KOS}, there is an important subtlety for
the resummed jet cross section related to the treatment of the invariant
mass of the jet. The structure of the large logarithmic corrections that 
are addressed by resummation depends on whether or not the jet is assumed 
to be massless at partonic threshold, even at the leading-logarithmic
(LL) level. This is perhaps 
surprising at first sight, because one might expect 
the jet mass to be generally inessential since it is typically much smaller 
than the jet's transverse momentum $p_T$ and in fact vanishes for 
lowest-order partonic scattering. However, the situation can be 
qualitatively understood as follows~\cite{KOS}: let us assume that we are 
defining the jet cross section from the total four-momentum deposited 
in a cone of aperture $R$~\footnote{Details of the jet definition 
do not matter for the present argument.}. Considering 
for simplicity the next-to-leading order, we can have contributions by 
virtual $2\to 2$ diagrams, or by $2\to 3$ real-emission diagrams. For 
the former, a single particle produces the (massless) jet, in case of 
the latter, there are configurations where two particles in the final 
state jointly form the jet. 

Then, for a jet forced to be massless at partonic threshold,
the contributions with two partons in the cone must either have 
one parton arbitrarily soft, or the two partons exactly collinear.
The singularities associated with these configurations cancel against 
analogous ones in 
the virtual diagrams, but because the constraint on the real-emission 
diagrams is so restrictive, large double- and single-logarithmic 
contributions remain after the cancellation. This will happen regardless
of the size $R$ of the cone aperture, implying that the coefficients of the
large logarithms will be independent of $R$. These final-state threshold
logarithms arising from the observed jet suppress the cross section near 
threshold. Their structure is identical to that of the threshold logarithms
generated by the recoiling ``jet'', because the latter is not observed
and is indeed massless at partonic threshold. The combined final-state
logarithms then act against the threshold logarithms associated with 
initial-state radiation which are positive and enhance the cross
section~\cite{dyresum}. 

If, on the other hand, the jet invariant mass is not constrained to vanish
near threshold, far more final states contribute-- in fact, there will
be an integration over the jet mass to an upper limit proportional
to the aperture of the jet cone. As the $2\to 3$ contributions are
therefore much less restricted, the cancellations of infrared
and collinear divergences between real and virtual diagrams leave 
behind only {\it single} logarithms~\cite{KOS}, associated with soft,
but not with collinear, emission. Compared to the previously discussed 
case, there is therefore no double-logarithmic suppression of the
cross section by the observed jet, and one expects the calculated
cross section to be larger. Also, the single-logarithmic terms 
will now depend on the jet cone size $R$.

The resummation for both these cases, with the jet massless or massive
at threshold, has been worked out in~\cite{KOS}. The study~\cite{KO}
of the resummed single-inclusive high-$p_T$ jet cross section assumed
massless jets at threshold. From a phenomenological point of view, 
however, we see no reason for demanding the jet to become massless 
at the partonic threshold. The experimental jet cross sections will,
at any given $p_T$, contain jet events with a large variety of jet 
invariant masses. NLO calculations of single-inclusive jet cross sections 
indeed reflect this: they have the property that jets produced at partonic 
threshold are integrated over a range of jet masses. This becomes evident 
in the available semi-analytical NLO 
calculations~\cite{Aversa:1988vb,Guillet,Jager:2002xm,ref:salesch}. For these, 
the jet cross 
section is obtained by assuming that the jet cone is relatively narrow,
in which case it is possible to treat the jet definition analytically,
so that collinear and infrared final-state divergences may be canceled
by hand. This approximation is referred to as the ``small-cone approximation
(SCA)''. Section~II.E in the recent calculation in~\cite{Jager:2002xm} 
explicitly demonstrates for the SCA that the threshold double-logarithms 
associated with the observed final-state jet cancel, as described above.

In light of this, we will study in this work the resummation in the more 
realistic case of jets that are massive at threshold. We will in fact make
use of the NLO calculation in the SCA approximation of~\cite{Jager:2002xm}
to ``match'' our resummed cross sections to finite (next-to-leading) order. 
Knowledge of analytical NLO expressions allows one to extract certain
hard-scattering coefficients that are finite at threshold and part
of the full resummation formula. These coefficients will be presented
and used in our paper for the first time.

We emphasize that the use of the SCA in our work is not to be regarded 
as a limitation to the usefulness of our results. First, the SCA is known 
to be very accurate numerically even at relatively large jet cone sizes 
of $R\sim 0.7$~\cite{Jager:2002xm,Guillet,ref:scavalid}. In addition, 
one may use our results to obtain ratios of 
the resummed over the NLO cross sections. Such ``$K$-factors'' are then 
expected to be extremely good approximations for the effects of higher orders 
even when one goes away from the SCA and uses, for example, a full NLO
Monte-Carlo integration code that allows to compute the jet cross section
for larger cone aperture and for other jet definitions (see, for example,
Ref.~\cite{Frixione:1997np}). We will therefore
in particular present $K$-factors for the resummed jet cross section in 
this paper. 

The paper is organized as follows: in Sec.~\ref{sec2} we provide 
the basic formulas for the single-inclusive-jet cross section 
at fixed order in perturbation theory, and discuss the SCA and the 
role of the threshold region. Section~\ref{sec3} presents details 
of the threshold resummation for the inclusive-jet cross section
and describes the matching to the analytical expressions for the 
NLO cross section in the SCA. In Sec.~\ref{sec4} we give phenomenological 
results for the Tevatron and for RHIC. 
Finally, we summarize our results in Sec.~\ref{sec5}. The Appendix 
collects the formulas for the hard-scattering coefficients in the 
threshold-resummed cross section mentioned above.

\section{Next-to-leading order single-inclusive jet cross 
section \label{sec2}}

Jets produced in high-energy hadronic scattering, $H_1(P_1) H_2(P_2) 
\rightarrow {\mathrm{jet}}(P_J) X$, are typically defined in terms of 
a deposition of transverse energy or four-momentum in a cone of aperture 
$R$ in pseudo-rapidity and azimuthal-angle space, 
with detailed algorithms specifying the jet kinematic variables in terms 
of those of the observed hadron momenta~\cite{Affolder:2001fa,Huth:1990mi,Ellis:1992qq,ref:othercone,ref:cluster,ref:discussion}. 
QCD factorization theorems allow 
to write the cross section for single-inclusive jet production in hadronic 
collisions in terms of convolutions of parton distribution functions 
with partonic hard-scattering functions~\cite{ref:fact}:
\begin{align}
\label{eq:parton}
d\sigma = \sum_{a,b}\, & \int dx_1 dx_2 \,
f_{a/H_1}\left(x_1,\mu_F^2\right) \, f_{b/H_2}\left(x_2,\mu_F^2\right) \,
d\hat{\sigma}_{ab}(x_1 P_1,x_2 P_2,P_J,\mu_F,\mu_R) \, ,
\end{align}
where the sum runs over all initial partons, quarks, anti-quarks, and 
gluons, and where $\mu_F$ and $\mu_R$ denote the factorization and 
renormalization scales, respectively. It is possible to use perturbation 
theory to describe the 
formation of a high-$p_T$ jet, as long as the definition of the jet
is infrared-safe. The jet is then constructed from a subset of 
the final-state partons in the short-distance reaction $ab\to 
{\mathrm{partons}}$, and a ``measurement function'' in the 
$d\hat{\sigma}_{ab}$ specifies the momentum $P_J$ of the jet 
in terms of the momenta of the final-state partons, in accordance
with the (experimental) jet definition. 

The computation of jet cross sections beyond the lowest order in 
perturbative QCD is rather complicated, due to the need for incorporating 
a jet definition and the ensuing complexity of the phase space, and due
to the large number of infrared singularities of soft and collinear origin 
at intermediate stages of the calculation. Different methods have been 
introduced that allow the calculation to be performed largely numerically
by Monte-Carlo ``parton generators'', with only the divergent terms 
treated in part analytically (see, for example, Ref.~\cite{Frixione:1997np}). 

A major simplification occurs if one assumes that the jet cone is rather 
narrow, a limit known as the ``small-cone approximation 
(SCA)''~\cite{Aversa:1988vb,Guillet,Jager:2002xm,ref:salesch}. In this
case, a semi-analytical computation of the NLO single-inclusive jet 
cross section can be performed, meaning that fully analytical expressions
for the partonic hard-scattering functions $d\hat{\sigma}_{ab}$ can 
be derived which only need to be integrated numerically against the parton 
distribution functions as shown in Eq.~(\ref{eq:parton}).
The SCA may be viewed as an expansion of the partonic cross section 
for small $\delta \equiv R/\cosh\eta$, where $\eta$ is the jet's
pseudo-rapidity. Technically, the parameter $\delta$ is the 
half-aperture of a geometrical cone around the jet axis, when 
the four-momentum of the jet is defined as simply the sum of the 
four-momenta of all the partons inside the 
cone~\cite{Aversa:1988vb,Jager:2002xm}. At small $\delta$, the
behavior of the jet cross section is of the form ${\cal A} \log(\delta) +  
{\cal B} + {\cal O}(\delta^2)$, with both ${\cal A}$ and ${\cal B}$
known from Refs.~\cite{Aversa:1988vb,Jager:2002xm}. Jet codes based on the SCA
have the virtue that they produce numerically stable results on 
much shorter time scales than Monte-Carlo codes. Moreover, as we shall 
see below, the relatively simple and explicit results for the NLO 
single-inclusive jet cross section obtained in the SCA are a great 
convenience for the implementation of threshold resummation, particularly 
for the matching needed to achieve full NLL accuracy.

\begin{figure}[htb]
\begin{center}
\epsfig{figure=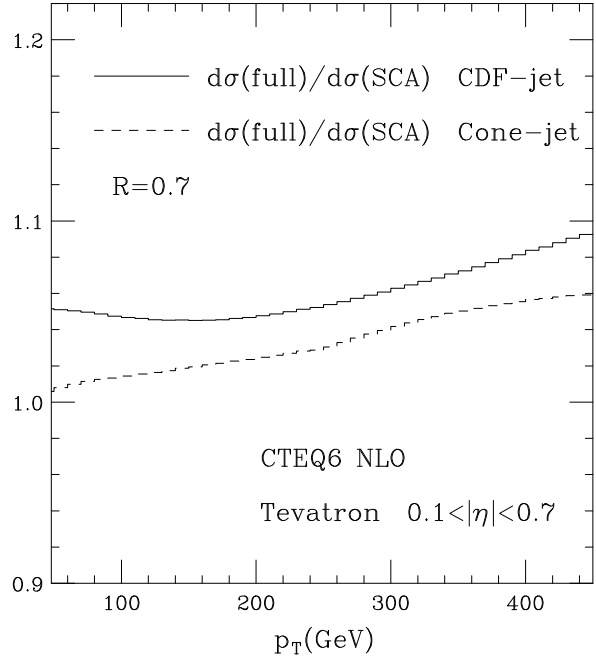,width=0.45\textwidth}
\end{center}
\vspace*{-.5cm}
\caption{Ratio between NLO jet cross sections at Tevatron 
at $\sqrt{S}=1800$~GeV, computed  
with a full Monte-Carlo code \cite{Frixione:1997np} and in the SCA. 
The solid line corresponds to the jet definition implemented by 
CDF~\cite{Affolder:2001fa} (with parameter $R_{sep}=1.3$), and the 
dashed one to the standard cone definition~\cite{Huth:1990mi}. In 
both cases the size of the jet cone is set to $R=0.7$, and the 
CTEQ6M NLO \cite{Pumplin:2002vw} parton distributions, evaluated 
at the factorization scale $\mu_F=P_T$, were used. 
\label{fig:1}}
\vspace*{.6cm}
\end{figure}

It turns out that the SCA is a very good approximation even for relatively
large cone sizes of up to 
$R\simeq 0.7$~\cite{Jager:2002xm,Guillet,ref:scavalid}, the 
value used by both Tevatron 
collaborations. Figure~\ref{fig:1} shows comparisons between the 
NLO cross sections for single-inclusive jet production obtained using 
a full Monte-Carlo code \cite{Frixione:1997np} and the SCA 
code of~\cite{Jager:2002xm}, for
$p\bar{p}$ collisions at c.m. energy $\sqrt{S}=1800$~GeV and very high $p_T$.
Throughout this paper we use the CTEQ6M~\cite{Pumplin:2002vw} NLO parton
distribution functions. We have chosen two different jet definitions 
in the Monte-Carlo calculation. One uses a conventional cone 
algorithm~\cite{Huth:1990mi}, the other the CDF jet 
definition~\cite{Affolder:2001fa}. One can see that the differences with 
respect to the SCA are of the order of only a few per cent. We note
that similar comparisons in the RHIC kinematic regime have been shown
in~\cite{Jager:2002xm}. In their recent paper~\cite{Abelev:2006uq}, 
the STAR collaboration used $R=0.4$, for which the SCA is even 
more accurate. 

Encouraged by this good agreement, we will directly use the SCA 
analytical results when performing the threshold resummation. As
stated in the Introduction, this is anyway not a limitation, because
we will also always provide the ratio of resummed over NLO cross sections
($K$-factors), which may then be used along with full NLO Monte-Carlo
calculations to obtain resummed cross sections for any desired 
cone size or jet algorithm.

A further simplification that we will make is to consider the 
cross section integrated over all pseudo-rapidities of the jet. 
As was discussed in~\cite{deFlorian:2005yj}, this considerably
reduces the complexity of the resummed expressions. By simply rescaling 
the resummed prediction by an appropriate ratio of NLO cross sections 
one can nonetheless obtain a very good approximation also for the 
resummation effects on the non-integrated cross section, at central 
rapidities~\cite{sv}. To perform the NLL threshold resummation for
the full rapidity-dependence of the jet cross section remains an
outstanding task for future work.

From Eq.~(\ref{eq:parton}), we find for the single-inclusive jet 
cross section integrated over all jet pseudo-rapidity $\eta$,
in the SCA:
\begin{align}
\label{eq:1}
\f{p_T^3\, d\sigma^{{\mathrm{SCA}}}(x_T)}{dp_T} = \sum_{a,b}\, &
\int_0^1 dx_1 \, f_{a/H_1}\left(x_1,\mu_F^2\right) \,
 \int_0^1 dx_2 \, f_{b/H_2}\left(x_2,\mu_F^2\right) \, \nn \\
  &\int_0^1 d\hat{x}_T \, \,  \delta\left(\hat{x}_T-\f{x_T}
{\sqrt{x_1 x_2}}\right)
   \, \int_{\hat{\eta}_{-}}^{\hat{\eta}_{+}} d\hat{\eta} \, 
\f{\hat{x}_T^4 \,s}{2} \,\f{d\hat{\sigma}_{ab}
(\hat{x}_T^2,\hat{\eta},R)}{d\hat{x}_T^2 d\hat{\eta}} \, ,
\end{align}
where as before $x_T\equiv 2 p_T/\sqrt{S}$ is the customary scaling 
variable, and $\hat{x}_T\equiv 2 p_T/\sqrt{s}$ with $s=x_1 x_2 S$ is 
its partonic counterpart. $\hat{\eta}$ is the partonic pseudo-rapidity, 
$\hat{\eta}=\eta-\frac{1}{2}\ln(x_1/x_2)$, which has the limits 
$\hat{\eta}_{+}=-\hat{\eta}_{-}=\ln[(1+\sqrt{1-\hat{x}_T^2})/\hat{x}_T]$.
The dependence of the partonic cross sections on $\mu_F$ and $\mu_R$ has
been suppressed for simplicity. The perturbative expansion of the 
$d\hat{\sigma}_{ab}$ in the coupling constant $\as(\mu_R)$ reads
\begin{align}
d\hat{\sigma}_{ab}(\hat{x}_T^2,\hat{\eta},R)
= &\as^2(\mu_R)\,\Big[ d\hat{\sigma}_{ab}^{(0)}(\hat{x}_T^2,\hat{\eta})+ 
\as(\mu_R)\,d\hat{\sigma}_{ab}^{(1)}(\hat{x}_T^2,\hat{\eta},R) + 
{\cal O}(\as^2) \Big] \, .
\end{align}
As indicated, the leading-order (LO) term $d\hat{\sigma}_{ab}$ 
has no dependence on the cone size $R$, because for this term a single 
parton produces the jet. The analytical expressions for the NLO terms 
$d\hat{\sigma}_{ab}^{(1)}$ have been obtained 
in~\cite{Aversa:1988vb,Jager:2002xm}. It is customary to express 
them in terms of a different set of variables, $v$ and $w$, that are
related to $\hat{x}_T$ and $\hat{\eta}$ by
\begin{equation}
\hat{x}_T^2=4 v w(1-v) \;\;, 
\;\;\;\;\;\; {\rm e}^{2\hat{\eta}}=\f{v w}{1-v} \,.
\end{equation}
Schematically, the NLO corrections to the partonic cross section 
for each scattering channel then take the form
\begin{eqnarray}
\label{eq:sigma1}
\f{s\, d\hat{\sigma}_{ab}^{(1)}(w,v,R)}{dw\,dv}
&=& A_{ab}(v,R)\, \delta(1-w) + B_{ab}(v,R)\, 
\left(\f{\ln(1-w)}{1-w} \right)_+ \nonumber \\
&&+C_{ab}(v,R)\, \left(\f{1}{1-w}  \right)_++ F_{ab}(w,v,R) \, ,
\end{eqnarray}
where the ``plus''-distributions are defined as usual by
\begin{equation}
\int_0^1 dw f(w)[g(w)]_+\equiv \int_0^1 dw (f(w)-f(1))g(w)\ ,
\end{equation} 
and where the $F_{ab}(w,v,\delta)$ collect all terms 
without distributions in $w$. Partonic threshold corresponds to the 
limit $w\to 1$. The ``plus''-distribution terms in Eq.~(\ref{eq:sigma1}) 
generate the large logarithmic corrections that are addressed by threshold 
resummation. At order $k$ of perturbation theory, the leading contributions 
are proportional to $\as^k \, \left(\f{\ln(1-w)}{1-w} \right)_+^{2k-1}$.
Performing the integration of these terms over $\hat{\eta}$, they turn
into contributions $\propto \as^k \ln^{2k}\left(1-\hat{x}_T^2\right)$,
as we anticipated in the Introduction, and as we shall show below. 
Subleading terms are down by one or more powers of the logarithm. 
We will now turn to the NLL resummation of the threshold logarithms.

\section{Resummed cross section \label{sec3}}

The resummation of the soft-gluon contributions is carried out in 
Mellin-$N$ moment space, where they 
exponentiate~\cite{dyresum,Catani:1996yz,KS,KOS,KOS1,LOS,BCMN}. 
At the same time,
in moment space the convolutions between the parton distributions 
and the partonic subprocess cross sections turn into ordinary products.
For our present calculation, the appropriate Mellin moments are in the 
scaling variable $x_T$~\footnote{We drop the superscript ``SCA'' from now on.}:
\begin{align}
\label{eq:moments}
\sigma(N)\equiv \int_0^1 dx_T^2 \, \left(x_T^2 \right)^{(N-1)} \f{p_T^3\, 
d\sigma(x_T)}{dp_T} \, .
\end{align}
In Mellin-$N$ space the QCD factorization formula in Eq.~(\ref{eq:1}) becomes
\begin{align}
\sigma(N)=\sum_{a,b} \,  f_{a/H1}^{N+1}(\mu_F^2) 
\,  f_{b/H2}^{N+1}(\mu_F^2) 
 \, \hat{\sigma}_{ab}(N)\, ,
\end{align}
where the $f^N_{a/H}$ are the moments of the parton distribution
functions,
\begin{align}
f^N_{a/H}(\mu_F^2) \equiv
\int_0^1 dx x^{N-1} f_{a/H}(x,\mu_F^2) \,,
\end{align}
and where
\begin{align}
 \hat{\sigma}_{ab}(N) \equiv
\f{1}{2} \int_0^1 dw  \int_0^1 dv\, \left[4 v(1-v)w\right]^{N+1}\,
 \f{s\, d\hat{\sigma}_{ab}(w,v)}{dw\,dv} \,.
\end{align}
The threshold limit $w\to 1$ corresponds to $N\to \infty$, and the LL
soft-gluon corrections contribute as $\as^m \ln^{2m}N$. The large-$N$ 
behavior of the NLO partonic cross sections can be easily obtained by using
\begin{align}
\int_0^1  \left[4 v(1-v)\right]^{N+1}\, f(v) = \int_0^1  
\left[4 v(1-v)\right]^{N+1}
\left( f\left(\f{1}{2}\right) +{\cal O}\left(\f{1}{N}\right) \right)\, .
\end{align}
Up to corrections suppressed by $1/N$ it is therefore possible to perform
the $v$-integration of the partonic cross sections by simply evaluating them 
at $v=1/2$. According to Eq.~(\ref{eq:sigma1}), when $v=1/2$ is combined 
with the threshold limit $w=1$, one has $\hat{x}_T=1$. It is worth 
mentioning that in the same limit one has $\hat{\eta}=0$, and therefore the 
coefficients for the soft-gluon resummation for the rapidity-integrated 
cross section agree with those for the cross section at vanishing partonic 
rapidity. This explains why generally the resummation for the 
rapidity-integrated hadronic cross section yields a good approximation to the 
resummation of the cross section integrated over only a finite
rapidity interval, as long as a region around $\eta=0$ is contained
in that interval~\cite{sv}.

\begin{figure}[t!]
\begin{center}
\vspace*{-0.2cm}
\epsfig{figure=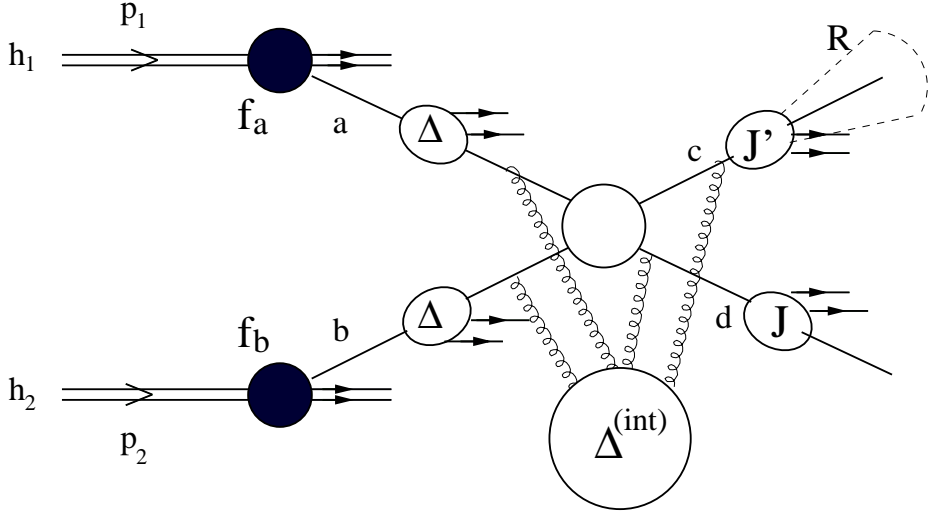,width=0.7\textwidth}
\end{center}
\vspace*{-.5cm}
\caption{Pictorial representation of the resummation formula 
in Eq.~(\ref{eq:res}).
\label{fig:grafico}}
\vspace*{.6cm}
\end{figure}

The resummation of the large logarithms in the partonic cross sections is 
achieved by showing that they exponentiate in Sudakov form factors.
The resummed cross section for each partonic subprocess is given by a 
formally rather simple expression in Mellin-$N$ space: 
\begin{align}
\label{eq:res}
\hat{\sigma}^{(res)}_{ab} (N)= \sum_{c,d} C_{ab}\,   
\Delta^{a}_N\, \Delta^{b}_N\,
J'^{c}_{N}\, J^{d}_N\, \left[ \sum_{I} G^{I}_{ab\to cd}\,
\Delta^{{\mathrm{(int)}} ab\rightarrow cd}_{I\, N}\right]
\, \hat{\sigma}^{{\mathrm{(Born)}}}_{ab\to cd} (N) \, ,
\end{align}
where the first sum runs over all possible final state partons 
$c$ and $d$, and the second over all possible color configurations 
$I$ of the hard scattering. Except for the Born cross sections
$\hat{\sigma}^{{\mathrm{(Born)}}}_{ab\to cd}$ (which we have
presented in earlier work~\cite{deFlorian:2005yj}), 
each of the $N$-dependent factors in Eq.~(\ref{eq:res}) is 
an exponential containing logarithms in $N$. The coefficients $C_{ab}$ 
collect all $N$-independent contributions, which partly arise from hard 
virtual corrections and can be extracted from comparison to the analytical 
expressions for the full NLO corrections in the SCA. Finally, the 
$G^{I}_{ab\to cd}$ are color weights obeying $\sum_I G^{I}_{ab\to cd}=1$.
The color interferences expressed by the sum over $I$ appear whenever
the number of partons involved in the process at Born level is 
larger than three, as it is the case here~\footnote{In a more general 
case without rapidity integration, the terms $G^{I}_{ab\to cd} \times 
\hat{\sigma}^{{\mathrm{(Born)}}}_{ab\to cd} (N)$ should be replaced by
the color-correlated Born cross sections $\hat{\sigma}^{
({\mathrm{Born}}\, I)}_{ab\to cd} (N)$~\cite{KOS,BCMN,KO}.}.
Figure~\ref{fig:grafico} gives a simple graphical mnemonic of the 
structure of the resummation formula, and of the origin of its
various factors, whose expressions we know present.

Effects of soft-gluon radiation collinear to the {\it initial-state} 
partons are exponentiated in the functions $\Delta^{a,b}_N$, which read:
\begin{equation}
\label{eq:delta}
\ln \Delta^{a}_N=  \int_0^1 \f{z^{N-1}-1}{1-z} \int_{\mu_F^2}^{(1-z)^2 Q^2} 
\f{dq^2}{q^2} A_a(\as(q^2))  \, ,
\end{equation}
(and likewise for $b$) where $Q^2=2 p_T^2$. $J^{d}_N$ is the exponent 
associated with collinear, both soft and hard, radiation in the unobserved
recoiling ``jet'',
\begin{equation}
\label{eq:j}
\ln J^{d}_N=  \int_0^1 \f{z^{N-1}-1}{1-z} \Big[ \int_{(1-z)^2 Q^2}^{(1-z) Q^2} 
\f{dq^2}{q^2} A_d(\as(q^2)) +\f{1}{2} B_d(\as((1-z)Q^2)) \Big]\, .
\end{equation}

The function $J'^{c}_{N}$ describes radiation in the observed
jet. As we reviewed in the Introduction, this function is 
sensitive to the assumption made about the jet's invariant 
mass at threshold~\cite{KOS} even at LL level. Our choice is to 
allow the jet to be massive at partonic threshold, which is consistent 
with the experimental definitions of jet cross sections and with the 
available NLO calculations. In this case, $J'^{c}_{N}$ is given as
\begin{equation}
\label{eq:jp}
\ln J'^{c}_N=  \int_0^1 \f{z^{N-1}-1}{1-z}\, C'_c(\as((1-z)^2Q^2)) 
\;\;\;\;\;\;\;\; {\mathrm{(jet \; massive \; at \; threshold)}}\, .
\end{equation}
A similar exponent was derived for this case in~\cite{KOS}. 
The expression given in Eq.(\ref{eq:jp}) agrees with the one 
of~\cite{KOS} to the required NLL accuracy. Notice that $J'^{c}_{N}$
contains only single logarithms, which arise from soft emission, 
whereas logarithms of collinear origin are absent. This is
explicitly seen in the NLO calculations in the SCA~\cite{Guillet,Jager:2002xm}
in which there is an integration over the jet mass up to a maximum value 
of ${\cal O}(\delta\sim R)$, even when the threshold limit is strictly 
reached. Collinear contributions that would usually generate large logarithms
are actually ``regularized'' by the cone size $\delta$ and give instead rise 
to $\log (\delta)$ terms in the perturbative cross sections. 
  
If, however, the jet is forced to be massless at partonic threshold the 
jet-function is identical to the function for an ``unobserved'' jet 
given in Eq.~(\ref{eq:j})~\cite{KOS,KO}:
\begin{equation}
\label{eq:jpmassless}
\ln J'^{c}_N= \ln J^{c}_N 
\;\;\;\;\;\;\;\; {\mathrm{(jet \; massless \; at \; threshold)}}  \, ,
\end{equation}
which produces a (negative) double-logarithm per emitted gluon,
because it also receives collinear contributions due to 
the stronger restriction on the gluon phase space. There is then
no dependence on $\log(\delta)$ in this case. 

Finally, large-angle soft-gluon emission is accounted for by the factor 
$\Delta^{{\mathrm{(int)}} ab\rightarrow cd}_{I\, N}$, which reads
\begin{equation}
\label{eq:interf}
\ln\Delta^{{\mathrm{(int)}} ab\rightarrow cd}_{I\, N} = 
 \int_0^1 \f{z^{N-1}-1}{1-z}\, D_{I\, ab\to cd}(\as((1-z)^2 Q^2))\, ,
 \end{equation}
and depends on the color configuration $I$ of the participating partons.

The coefficients $A_a,\, B_a,\,C'_a,\, D_{I\, ab\to cd}$ in 
Eqs.~(\ref{eq:delta}),(\ref{eq:j}),(\ref{eq:jp}),(\ref{eq:interf})
are free of large logarithmic contributions and are given as
perturbative series in the coupling constant $\as$:
\begin{equation}
\label{eq:series}
{\cal F}(\as)=\frac{\as}{\pi}{\cal F}^{(1)}+\left(\frac{\as}{\pi}\right)^2
{\cal F}^{(2)}+\ldots
\end{equation}
for each of them. For the resummation to NLL accuracy we need the 
coefficients $A_a^{(1)}$, $A_a^{(2)}$, $B_a^{(1)}$, $C'^{(1)}_a$, 
and $D_{I\, ab\to cd}^{(1)}$. The last of these depends on the specifics 
of the partonic process under consideration; all the others are universal 
in the sense that they only distinguish whether the parton they
are associated with is a quark or a gluon. The LL and NLL coefficients 
$A_a^{(1)}$, $A_a^{(2)}$ and  $B_a^{(1)}$ are well known~\cite{KT}:
\begin{equation} 
\label{A12coef} 
A^{(1)}= C_a
\;,\;\;\;\; A^{(2)}=\frac{1}{2} \; C_a K \;,\;\;\;\; B_a^{(1)}=\gamma_a
\end{equation} 
with
\begin{equation} 
\label{kcoef} 
K = C_A \left( \frac{67}{18} - \frac{\pi^2}{6} \right)  
- \frac{5}{9} N_f \;, 
\end{equation}
where $C_g=C_A=N_c=3$, $C_q=C_F=(N_c^2-1)/2N_c=4/3$, $\gamma_q=-3/2 C_F=-2$,
$\gamma_g=-2\pi \b0$, and $N_f$ is the number of flavors.
The coefficients $C'^{(1)}_a$ needed in the case of jets that are massive
at threshold, may be obtained by comparing the first-order expansion of the 
resummed formula to the analytic NLO results in the SCA. They are also
universal and read
\begin{equation} 
\label{C'coef} 
C'^{(1)}_a= - C_a \log\left( \f{\delta^2 }{8} \right) \, .
\end{equation}
This coefficient contains the anticipated dependence on $\log(\delta)$ 
that regularizes the final-state collinear configurations. As expected
from a term of collinear origin, the exponent~(\ref{eq:jp}) hence 
provides one power of $\log(\delta)$ for each perturbative order.

The coefficients $D_{I\, ab \to c d}^{(1)}$ governing the 
exponentiation of large-angle soft-gluon emission to NLL accuracy, 
and the corresponding ``color weights'' $G_{I\, ab \to c d}$,
depend both on the partonic process and on its ``color configuration''. 
They can be obtained from~\cite{KOS,KOS1,KO,kidonakis,mert} where the soft 
anomalous dimension matrices were computed for all partonic processes. 
The results are given for the general case of arbitrary
partonic rapidity $\hat{\eta}$. As discussed above, the coefficients 
for the rapidity-integrated cross section may be obtained by setting 
$\hat{\eta}=0$. We have presented the full set of the 
$D_{I\, ab \to c d}^{(1)}$ and $G_{I\, ab \to c d}$ in the Appendix 
of our previous paper~\cite{deFlorian:2005yj}. 

Before we continue, we mention that one expects that a jet
cross section defined by a cone algorithm will also have so-called
``non-global'' threshold logarithms~\cite{dg1,bks}. These logarithms
arise when the observable is sensitive to radiation in only a limited
part of phase space, as is the case in presence of a jet cone. For
instance, a soft gluon radiated at an angle {\it outside} 
the jet cone may itself emit a secondary gluon at large angle that happens
to fall {\it inside} the jet cone, thereby becoming part of the 
jet~\cite{dg1,bks}. Such configurations appear first at the 
next-to-next-to-leading order, but may produce threshold logarithms 
at the NLL level. One may therefore wonder if our NLL resummation
formulas given above are complete. Fortunately, as an explicit example
in~\cite{dg1} shows, it turns out that these effects are suppressed
as $R\log(R)$ in the SCA. They may therefore be neglected at the
level of approximation we are making here. Given their only mild
suppression as $R\to 0$, a study of non-global logarithms
in hadronic jet cross sections is an interesting topic for future
work. 

Returning to our resummed formulas, it is instructive to consider 
the structure of the leading logarithms.
The LL expressions for the radiative factors are
\begin{eqnarray}\label{DJfct1}
\Delta^a_N&=&  \exp\left[ \frac{\as}{\pi}C_a\ln^2 (N) \right] \; , \nn \\
J^d_N&=& \exp\left[ -\frac{\as}{2\pi}C_d \ln^2 (N) \right]  \; .
\end{eqnarray}
As discussed above, $J'^c$ does not contribute at the double-logarithmic 
level. Therefore, for a given partonic channel, the leading logarithms are
\begin{align}
\label{eq:res1}
\hat{\sigma}^{{\rm (res)}}_{ab\to cd} (N) \propto 
\exp\left[ \frac{\as}{\pi}\left(C_a+C_b-\frac{1}{2} C_d\right)
\ln^2 (N) \right] \; .
\end{align}
The exponent is positive for each partonic channel, implying that the 
soft-gluon effects will increase the cross section. This enhancement arises 
from the initial-state radiation represented by the $\Delta_{a,b}$ and  
is related to the fact that finite partonic cross sections are obtained 
after collinear (mass) factorization~\cite{dyresum,Catani:1996yz}. In the 
$\overline{{\mathrm{MS}}}$ scheme such an enhancing contribution 
is (for a given parton species) twice as large as the suppressing one 
associated with final-state radiation in $J^d$, for which no mass 
factorization is needed. For quark or anti-quark initiated processes, 
the color factor combination appearing in Eq.~(\ref{eq:res1}) ranges
from $2 C_F-C_F/2=2$ for the  $qq\rightarrow qq$  channel to 
$2 C_F-C_A/2=7/6$ for $q\bar{q}\rightarrow gg$, while for those involving 
a quark-gluon initial state one has larger factors, 
$C_F+C_A-C_F/2=11/3$ (for $qg\to qg$) or $C_F+C_A-C_A/2=17/6$ 
(for $qg\to gq$). Yet larger factors are obtained for gluon-gluon
scattering, with $2 C_A-C_A/2=9/2$ for $gg\rightarrow gg$ and 
$2 C_A-C_F/2=16/3$ for $gg \rightarrow q\bar{q}$. Initial states 
with more gluons therefore are expected to receive the larger
resummation effects. We mention that if the observed jet is assumed 
strictly massless at threshold, an extra suppression term proportional 
to $J^c$ arises (see Eq.~(\ref{eq:jpmassless})). 

It is customary to give the NLL expansions of the Sudakov exponents in
the following way~\cite{Catani:1996yz}:
\beeq
\label{lndeltams}
\!\!\! \!\!\! \!\!\! \!\!\! \!\!\!
\ln \Delta_N^a(\as(\mu_R^2),Q^2/\mu_R^2;Q^2/\mu_F^2) 
&\!\!=\!\!& \ln N \;h_a^{(1)}(\lambda) +
h_a^{(2)}(\lambda,Q^2/\mu_R^2;Q^2/\mu_F^2) + 
{\cal O}\left(\as(\as \ln N)^k\right) \,,\\
\label{lnjfun}
\ln J_N^a(\as(\mu_R^2),Q^2/\mu_R^2) &\!\!=\!\!& \ln N \;f_a^{(1)}(\lambda) +
f_a^{(2)}(\lambda,Q^2/\mu_R^2) + {\cal O}\left(\as(\as \ln N)^k\right) \,,\\
\ln J'^{a}_N(\as(\mu_R^2)) &\!\!=\!\!& \frac{C'^{(1)}_a}{2\pi b_0} \;
\ln (1-2\lambda) + {\cal O}\left(\as(\as \ln N)^k\right) \,,\\
\ln\Delta^{{\mathrm{(int)}} ab\rightarrow cd}_{I\, N}(\as(\mu_R^2))
&\!\!=\!\!& \frac{D_{I\, ab \to c d}^{(1)}}{2\pi b_0} \;\ln (1-2\lambda) +
{\cal O}\left(\as(\as \ln N)^k\right) \,,
\eeeq
with $\lambda=\b0 \as(\mu^2_R) \ln N$. The LL and NLL auxiliary 
functions $h_a^{(1,2)}$ and $f_a^{(1,2)}$ are
\begin{align} 
\label{h1fun}
h_a^{(1)}(\la) =&+ \f{A_a^{(1)}}{2\pi \b0 \la} 
\left[ 2 \la+(1-2 \la)\ln(1-2\la)\right] \;,\\ 
h_a^{(2)}(\la,Q^2/\mu^2_R;Q^2/\mu_{F}^2) 
=&-\f{A_a^{(2)}}{2\pi^2 \b0^2 } \left[ 2 \la+\ln(1-2\la)\right] - 
\f{A_a^{(1)} \gamma_E}{\pi \b0 } \ln(1-2\la)\nn \\ 
&+ \f{A_a^{(1)} \bone}{2\pi \b0^3} 
\left[2 \la+\ln(1-2\la)+\f{1}{2} \ln^2(1-2\la)\right]\nn \\ 
\label{h2fun}
&+ \f{A_a^{(1)}}{2\pi \b0}\left[2 \la+\ln(1-2\la) \right]  
\ln\f{Q^2}{\mu^2_R}-\f{A_a^{(1)}}{\pi \b0} \,\la \ln\f{Q^2}{\mu^2_{F}} \;,  
\end{align} 
\beeq
\label{fll}
f_a^{(1)}(\lambda) =
&-&\frac{A_a^{(1)}}{2\pi b_0 \lambda}\Bigl[(1-2\lambda)
\ln(1-2\lambda)-2(1-\lambda)
\ln(1-\lambda)\Bigr] \; , \\
\label{fnll}
f_a^{(2)}(\lambda,Q^2/\mu^2_R) =
&-&\frac{A_a^{(1)} b_1}{2\pi b_0^3}\Bigl[\ln(1-2\lambda)
-2\ln(1-\lambda)+\f{1}{2}\ln^2(1-2\lambda)-\ln^2(1-\lambda)\Bigr] \nonumber \\
&+&\frac{B_a^{(1)}}{2\pi b_0}\ln(1-\lambda)
-\frac{A_a^{(1)}\GE}{\pi b_0}\Bigl[\ln(1-\lambda)
-\ln(1-2\lambda)\Bigr] \\
&-&\frac{A_a^{(2)}}{2\pi^2 b_0^2}\Bigl[2\ln(1-\lambda)
-\ln(1-2\lambda)\Bigr] 
+ \frac{A_a^{(1)}}{2\pi b_0}\Bigl[2\ln(1-\lambda)
-\ln(1-2\lambda)\Bigr] \ln\frac{Q^2}{\mu^2_R} \; , \nonumber
\eeeq
where
$\b0, \bone$ are the first two coefficients of the QCD 
$\beta$-function:
\begin{align}
\b0 &= \frac{1}{12 \pi} \left( 11 C_A - 2 N_f \right) \; ,
\quad\quad \bone=  \frac{1}{24 \pi^2} 
\left( 17 C_A^2 - 5 C_A N_f - 3 C_F N_f \right) \; .
\label{bcoef}
\end{align}

The $N$-independent coefficients $C_{ab}$ in Eq.~(\ref{eq:res}), 
which include the hard virtual corrections, have the perturbative
expansion
\begin{equation} 
C_{ab}=1+\f{\as}{\pi}\, C_{ab}^{(1)} + {\cal O}(\as^2) \, .
\end{equation} 
The $C_{ab}^{(1)}$ we need to NLL are obtained by comparing the 
${\cal O}(\as)$-expansion (not counting the overall factor $\as^2$ of
the Born cross sections) of the resummed expression with the fixed-order 
NLO result for the process, as given in~\cite{Aversa:1988vb,Jager:2002xm}. 
The full analytic expressions for the $C_{ab}^{(1)}$ are rather lengthy
and will not be given here. For convenience, we present them in
numerical form in the Appendix. We note that apart from being
useful for extracting the coefficients $C_a'^{(1)}$ and $C_{ab}^{(1)}$,
the comparison of the ${\cal O}(\as)$-expanded resummed result 
with the full NLO cross section also provides an excellent check 
of the resummation formula, since one can verify that all leading
and next-to-leading logarithms are properly accounted for by 
Eq.~(\ref{eq:res}).  

The improved resummed hadronic cross section is finally obtained by 
performing an inverse Mellin transformation, and by properly matching 
to the NLO cross section $p_T^3\, d\sigma^{\rm (NLO)}(x_T)/dp_T$ as follows:
\begin{align}
\label{hadnres}
\f{p_T^3\, d\sigma^{\rm (match)}(x_T)}{dp_T} &= \sum_{a,b,c}\,
\;\int_{C_{MP}-i\infty}^{C_{MP}+i\infty}
\;\frac{dN}{2\pi i} \;\left( x_T^2 \right)^{-N+1}
\; f_{a/H_1}^N(\mu_F^2) \; f_{b/H_2}^N(\mu_F^2) \;  
 \nn \\
&\times \left[ \;
\hat{\sigma}^{\rm (res)}_{ab\to cd} (N)
- \left( \hat{\sigma}^{{\mathrm{(res)}}}_{ab\to cd} (N)
\right)_{\rm (NLO)} \, \right] 
+\f{p_T^3\, d\sigma^{\rm (NLO)}(x_T)}{dp_T} 
 \;\;,
\end{align}
where $\hat{\sigma}^{{\mathrm{(res)}}}_{ab\to cd}$ is given 
in Eq.~(\ref{eq:res}) and 
$(\hat{\sigma}^{{\mathrm{(res)}}}_{ab\to cd})_{\rm (NLO)}$
represents its perturbative truncation at NLO. Thus, as a result 
of this matching procedure, in the final cross section in 
Eq.~(\ref{hadnres}) the NLO cross section is exactly taken into
account, and NLL soft-gluon effects are resummed beyond those 
already contained in the NLO cross section.

The  functions $h_a^{(1,2)}(\lambda)$ and $f_a^{(1,2)}(\lambda)$
in Eqs.~(\ref{h1fun})-(\ref{fnll}) 
are singular at the points $\lambda=1/2$ and/or $\lambda=1$.
These singularities are related to the divergent behavior of the 
perturbative running coupling  $\as$ near the Landau pole, and we deal with 
them by using the {\em Minimal Prescription} introduced in  
Ref.~\cite{Catani:1996yz}. In the evaluation of the inverse Mellin 
transformation in Eq.~(\ref{hadnres}), the constant $C_{MP}$ is chosen 
in such a way that all singularities in the integrand are to the left 
of the integration contour, except for the Landau singularities, that 
are taken to lie to its far right. We note that an alternative to such 
a definition one could choose to expand the resummed formula to a finite 
order, say, next-to-next-to-leading order (NNLO), and neglect all terms of
yet higher order. This approach was adopted in Ref.~\cite{KO}. We
prefer to keep the full resummed formula in our phenomenological 
applications since, depending on kinematics, high orders in perturbation
theory may still be very relevant~\cite{deFlorian:2005yj}. It was
actually shown in~\cite{Catani:1996yz} that the results obtained within
the Minimal Prescription converge asymptotically to the perturbative series.

This completes the presentation of all ingredients to the NLL threshold 
resummation of the hadronic single-inclusive jet cross section. We will
now turn to some phenomenological applications.

\section{Phenomenological Results \label{sec4}}
We will study the effects of threshold resummation on the single-inclusive 
jet cross section in $p\bar{p}$ collisions at $\sqrt{S}=1.8$~TeV and 
$\sqrt{S}=630$~GeV c.m. energies, and in $pp$ collisions at 
$\sqrt{S}=200$~GeV. These choices are relevant for comparisons 
to Tevatron and RHIC data, respectively. Unless otherwise stated, we 
always set the factorization and renormalization scales to $\mu_F=\mu_R=p_T$ 
and use the NLO CTEQ6M~\cite{Pumplin:2002vw} set of parton distributions, 
along with the two-loop expression for the strong coupling constant 
$\as$. 

\begin{figure}[t!]
\begin{center}
\vspace*{-0.6cm}
\epsfig{figure=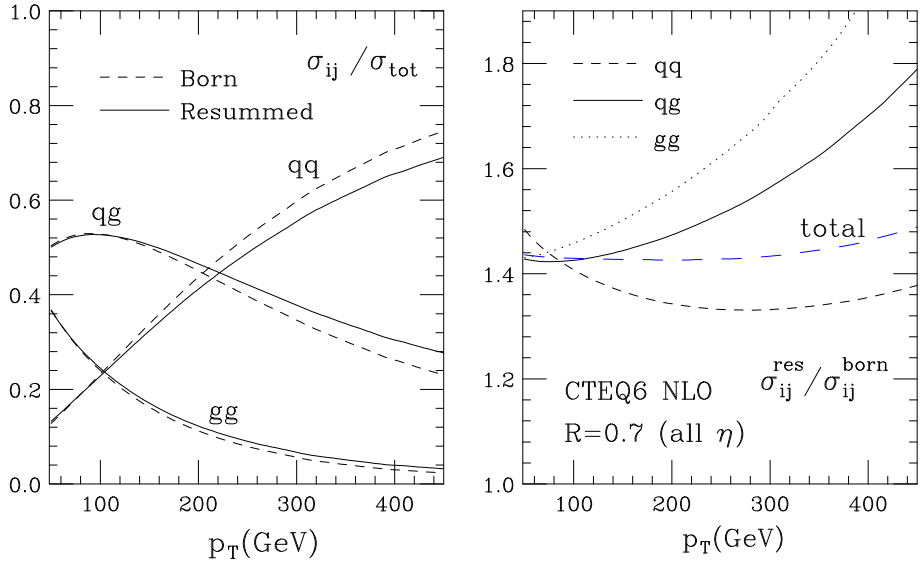,width=0.7\textwidth}
\end{center}
\vspace*{-.5cm}
\caption{Left: relative contributions of the various partonic initial 
states to the single-inclusive jet cross section in $p\bar{p}$ collisions
at $\sqrt{S}=1.8$~TeV, at Born level (dashed) and for the NLL resummed
case (solid). We have chosen the jet cone size $R=0.7$. Right: ratios 
between resummed and Born contributions for the various channels, and 
for the full jet cross section.
\label{fig:ratio}}
\vspace*{.6cm}
\end{figure}

We will first analyze the relevance of the different subprocesses 
contributing to single-jet production. The left part of Fig.~\ref{fig:ratio} 
shows the relative contributions by ``$qq$'' ($qq$, $qq'$, $q\bar{q}$ and 
$q\bar{q}'$ combined), $qg$ and $gg$ initial states at Born level (dashed
lines) and for the NLL resummed case (without matching, solid lines). Here we 
have chosen the case of $p\bar{p}$ collisions at $\sqrt{S}=1.8$~TeV. As
can be seen, the overall change in the curves when going from Born level
to the resummed case is moderate. The main noticeable effect is an increase
of the relative importance of processes with gluon initial states
toward higher $p_T$, compensated by a similar decrease in that of 
the $qq$ channels. In the right part of Fig.~\ref{fig:ratio} we show the
enhancements from threshold resummation for each initial partonic 
state individually, and also for their sum. At the higher $p_T$, where
threshold resummation is expected to be best applicable, the enhancements
are biggest for the $gg$ channel, followed by the $qg$ one. All patterns 
observed in Fig.~\ref{fig:ratio} are straightforwardly understood 
from Eq.~(\ref{eq:res1}), which demonstrates that resummation yields 
bigger enhancements when the number of gluons in the initial state
is larger. We note that results very similar to those shown in the 
figure hold also at $\sqrt{S}=630$~GeV, if the same value of 
$x_T=2p_T/\sqrt{S}$ is considered. This remains qualitatively true 
even when we go to $pp$ collisions at $\sqrt{S}=200$~GeV, except for the
 larger enhancement in the quark contribution, due to the 
dominance of the $qq$ channel (instead of the $q\bar{q}$ as in $p\bar{p}$ collisions)
with a larger color factor combination in the Sudakov exponent 
(see the discussion after Eq.~(\ref{DJfct1})). 

\begin{figure}[t!]
\begin{center}
\vspace*{-0.1cm}
\epsfig{figure=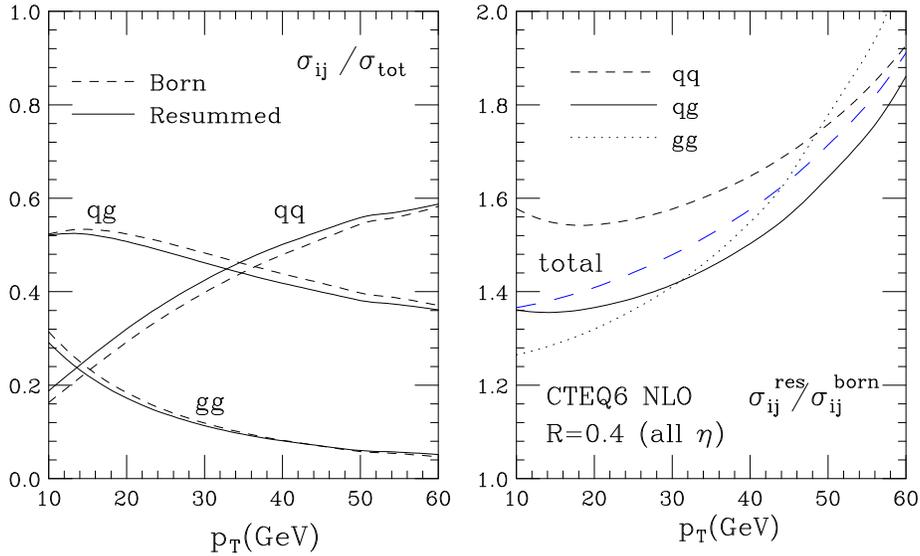,width=0.7\textwidth}
\end{center}
\vspace*{-.5cm}
\caption{Same as Fig.~\ref{fig:ratio}, but for $pp$ collisions at
$\sqrt{S}=200$~GeV and $R=0.4$.
\label{fig:ratio1}}
\vspace*{.6cm}
\end{figure}

Figure~\ref{fig:ratio1} repeats the studies made for Fig.~\ref{fig:ratio}
for this case. As one can see, if the same $x_T$ as in Fig.~\ref{fig:ratio}
is considered, the $qq$ scattering contributions are overall slightly less 
important. At the same time, resummation effects are overall somewhat
larger because the $p_T$ values are now much smaller than in 
Fig.~\ref{fig:ratio}, so that the strong coupling constant that appears
in the resummed exponents is larger.

\begin{figure}[t!]
\begin{center}
\vspace*{-0.1cm}
\epsfig{figure=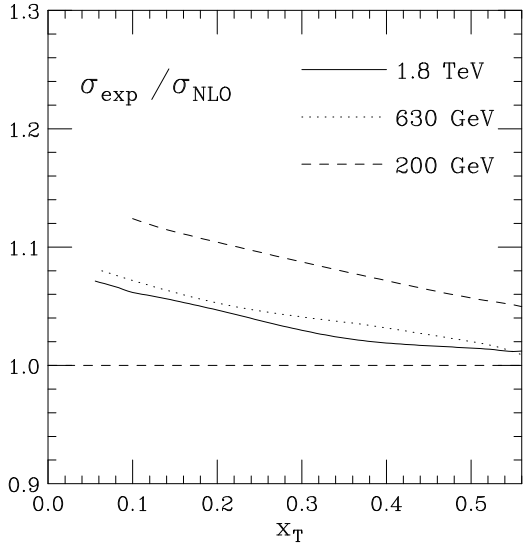,width=0.4\textwidth}
\end{center}
\vspace*{-.5cm}
\caption{Ratio between the expansion to NLO of the (unmatched) 
resummed cross section and the full NLO one (in the SCA),
for $p\bar{p}$ collisions at $\sqrt{s}=1.8$ TeV (solid) and 
$\sqrt{s}=630$ GeV (dots), and for $pp$ collisions at $\sqrt{s}=200$~GeV
(dashed). 
\label{fig:soft}}
\vspace*{.6cm}
\end{figure}

Before presenting the results for the matched NLL resummed jet cross section
and $K$-factors, 
we would like to identify the kinematical region where the logarithmic terms  
constitute the bulk of the perturbative corrections and subleading 
contributions are unimportant. Only in these is the resummation expected 
to provide an accurate picture of the higher-order terms. We can determine
this region by comparing the resummed formula, expanded to NLO, to the 
full fixed-order (NLO) perturbative result, that is, by comparing 
the last two terms in Eq.~(\ref{hadnres}). Figure~\ref{fig:soft} shows 
this comparison for both Tevatron energies and for the RHIC case, as 
function of the ``scaling'' variable $x_T$. As can be observed, the 
expansion correctly reproduces the NLO result within at most a few
per cent over a region corresponding to $p_T\gtrsim 200$~GeV for the
higher Tevatron energy, and to $p_T\gtrsim 30$~GeV at RHIC.
This demonstrates that, at this order, the perturbative 
corrections are strongly dominated by the terms of soft and/or collinear 
origin that are addressed by resummation. The accuracy of the expansion 
improves toward the larger values of the jet
transverse momentum, were one approaches the threshold limit more closely.

\begin{figure}[htb]
\begin{center}
\vspace*{0.4cm}
\epsfig{figure=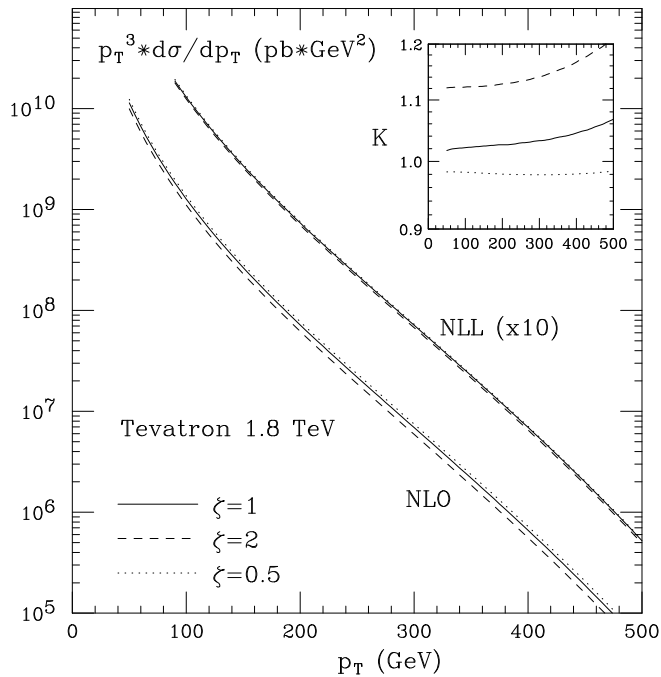,width=0.5\textwidth}
\end{center}
\vspace*{-.5cm}
\caption{NLO and NLL results for the single-inclusive jet cross section in
$p\bar{p}$ collisions at $\sqrt{S}=1.8$ TeV, for different values of the 
renormalization and factorization scales. We have chosen $R=0.7$. 
The inset plot shows the 
corresponding $K$-factors as defined in Eq.~(\ref{eq:kres}).
\label{fig:a3}}
\vspace*{.6cm}
\end{figure}

\begin{figure}[htb]
\begin{center}
\vspace*{-0.6cm}
\epsfig{figure=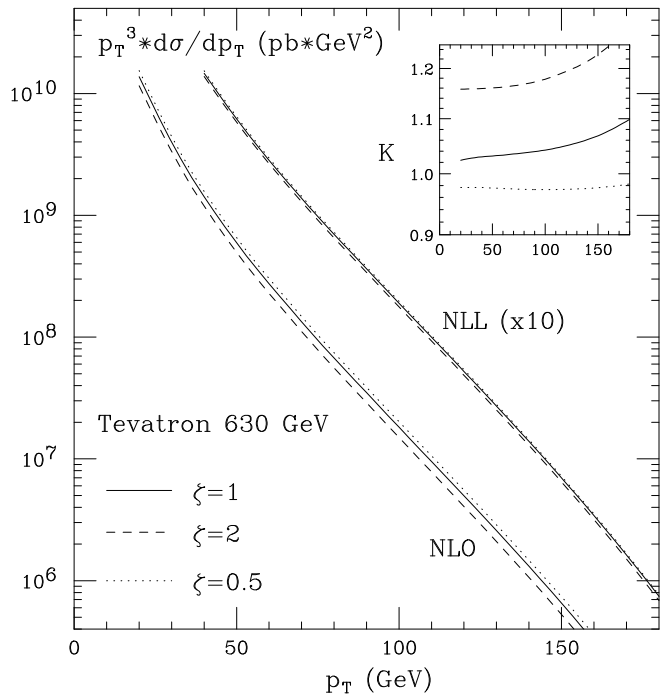,width=0.5\textwidth}
\end{center}
\vspace*{-.5cm}
\caption{Same as Fig.~\ref{fig:a3}, but for $\sqrt{S}=$630~GeV.
\label{fig:a4}}
\vspace*{.6cm}
\end{figure}

\begin{figure}[htb]
\begin{center}
\vspace*{-0.6cm}
\epsfig{figure=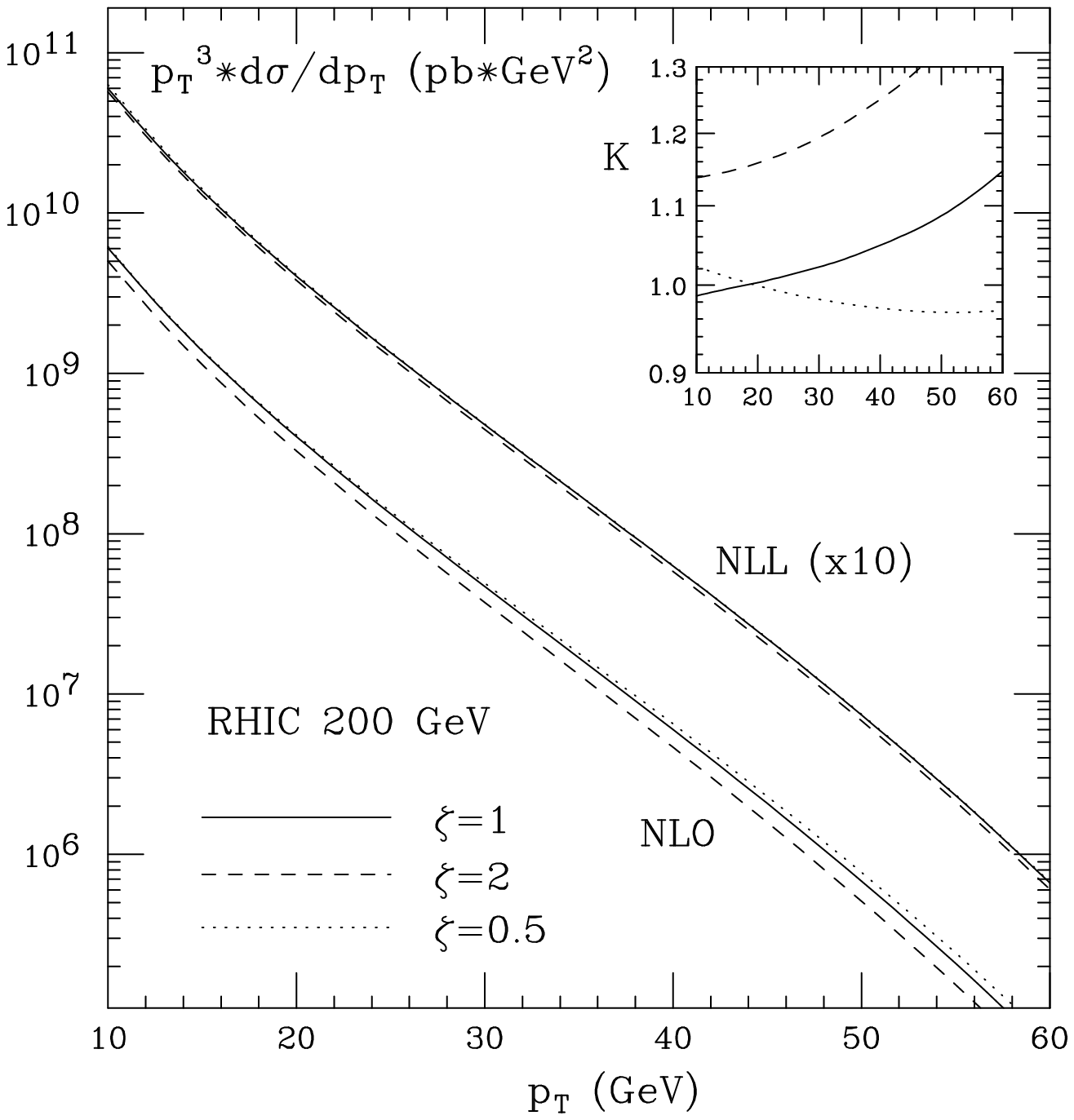,width=0.5\textwidth}
\end{center}
\vspace*{-.5cm}
\caption{Same as Fig.~\ref{fig:a3}, but for $pp$ collisions at 
$\sqrt{S}=$200~GeV and $R=0.4$.
\label{fig:a4a}}
\vspace*{.6cm}
\end{figure}

Having established the importance of the threshold corrections in a 
kinematic regime of interest for phenomenology, we show in 
Fig.~\ref{fig:a3} the impact of the resummation on the 
predicted single-jet cross section at $\sqrt{S}=1.8$ TeV. NLO and NLL 
resummed results are presented, computed at three different 
values of the factorization and renormalization scales, defined by 
$\mu_F=\mu_R=\zeta p_T$ (with $\zeta=1, 2, 1/2$). The most noticeable 
effect is a remarkable reduction in the scale dependence of the cross 
section. This observation was also made in the previous study~\cite{KO}.
If, as customary, one defines a theoretical scale ``uncertainty'' 
by $\Delta\equiv (\sigma(\zeta=0.5)-\sigma(\zeta=2))/ \sigma(\zeta=1)$, the 
improvement is considerable. While $\Delta$ lies between 20 and 25$\%$ 
at NLO, it never exceeds 8$\%$ for the matched NLL result. The inset plot 
shows the NLL $K$-factor, defined as
\begin{equation}
\label{eq:kres}
K^{{\rm (res)}} = \f{{d\sigma^{\rm (match)}}/{dp_T}}
{{d\sigma^{\rm (NLO)}}/{dp_T}}\, ,
\end{equation}
at each of the scales. The corrections from resummation on top of NLO
are typically very moderate, at the order of a few per cent, depending 
on the set of scales chosen. The higher-order corrections increase for 
larger values of the jet transverse momentum. These findings are again 
consistent with those of~\cite{KO}, even though more detailed comparisons 
reveal some quantitative differences that must be related to either
the different choice of the resummed final-state jet function in~\cite{KO}
(see discussion in Sec.~\ref{sec3}), or to the fact that~\cite{KO} uses
only a NNLO expansion of the resummed cross section. The main features of
our results remain unchanged when we go to the Tevatron-run~II energy
of $\sqrt{S}=1.96$~TeV, at which measured jet cross sections are now 
available~\cite{tevrun2}. Quantitatively very similar results are also 
found for the lower Tevatron center-of-mass energy, as seen in 
Fig.~\ref{fig:a4}. In the case of $pp$ collisions at $\sqrt{s}=200$~GeV,
presented in Fig.~\ref{fig:a4a}, a similar pattern emerges, even though 
the resummation effects tend to be overall somewhat more substantial here. 


\begin{figure}[htb]
\begin{center}
\vspace*{-0.6cm}
\epsfig{figure=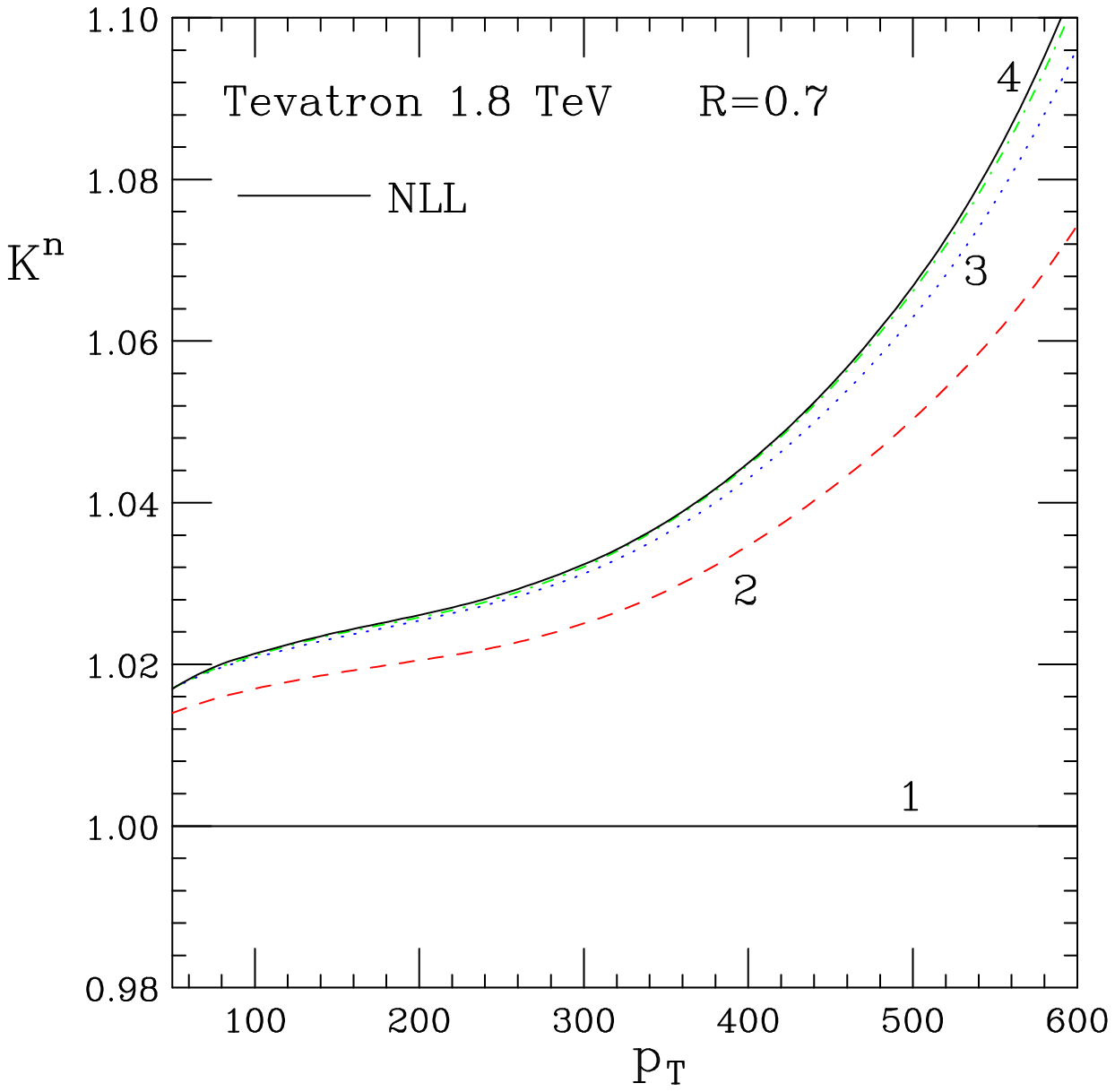,width=0.5\textwidth}
\end{center}
\vspace*{-.5cm}
\caption{Soft-gluon $K^n$ factors as defined in Eq.~(\ref{ksoftg}),
for $p\bar{p}$ collisions at $\sqrt{S}=1.8$~TeV.
\label{fig:expan}}
\vspace*{.6cm}
\end{figure}

In Fig.~\ref{fig:expan} we analyze how the resummation 
effects build up order by order in perturbation theory. We 
expand the matched resummed formula beyond NLO and define the ``partial''
soft-gluon $K$-factors as
\begin{equation} \label{ksoftg}
K^n\;\equiv\; \f{{\left. d\sigma^{\rm{(match)}}/{dp_T}\right|_{{\cal O}
(\as^{2+n})}}}{{d\sigma^{\rm (NLO)}}/{dp_T}} \; ,
\end{equation}
which for $n=2,3,\ldots$ give the additional enhancement over full NLO due 
to the ${\cal O}(\as^{2+n})$ terms in the resummed formula\footnote{We
recall that the Born cross sections are of ${\cal O}(\as^2)$, hence
the additional power of two in this definition.}. Formally,
$K^1=1$ and $K^{\infty}=K^{{\rm (res)}}$ of Eq.~(\ref{eq:kres}). The 
results for $K^{2,3,4,\infty}$ are given in the figure, 
for the case of $p\bar{p}$ collisions at $\sqrt{S}=1.8$~TeV. One can see 
that contributions beyond N$^3$LO ($n=3$) are very small, and that 
the ${\cal O}(\as^{6})$ result can hardly be distinguished from the 
full NLL one. 

It is interesting to contrast the rather modest enhancement of the 
jet cross section by resummation to the dramatic resummation effects 
that we observed in~\cite{deFlorian:2005yj} in the case of single-inclusive 
pion production, $H_1 H_2\to \pi X$, in fixed-target scattering at typical
c.m. energies of $\sqrt{S}\sim 30$~GeV. Even though in both cases the 
same partonic processes are involved at Born level, there are several 
important differences. First of all, the values of $p_T$ are much 
smaller in fixed-target scattering (even though roughly similar values
of $x_T=2p_T/\sqrt{S}$ are probed), so that the strong coupling
constant $\as(p_T)$ is larger and resummation effects are bound to be more
significant. Furthermore, for the process
$H_1 H_2\to \pi X$ one needs to introduce
fragmentation functions into the theoretical calculation that describe
the formation of the observed hadron from a final-state parton. As 
the hadron takes only a certain fraction $z\gtrsim 0.5$ of the parent 
parton's momentum, the partonic hard-scattering necessarily has to be
at the higher transverse momentum $p_T/z$ in order to produce
a hadron with $p_T$. Thus one is closer to partonic threshold than
in case of a jet produced with transverse momentum $p_T$ which takes
{\it all} of a final-state parton's momentum. In addition, it turns 
out~\cite{deFlorian:2005yj,Cacciari:2001cw} that due to the factorization of 
final-state collinear singularities associated with the fragmentation 
functions, the ``jet'' function $J'^{c}_{N}$ in the resummation formula 
Eq.~(\ref{eq:res}) is to be replaced by a factor $\Delta_N^c$, which 
has enhancing double logarithms. 

\begin{figure}[htb]
\begin{center}
\vspace*{-0.6cm}
\epsfig{figure=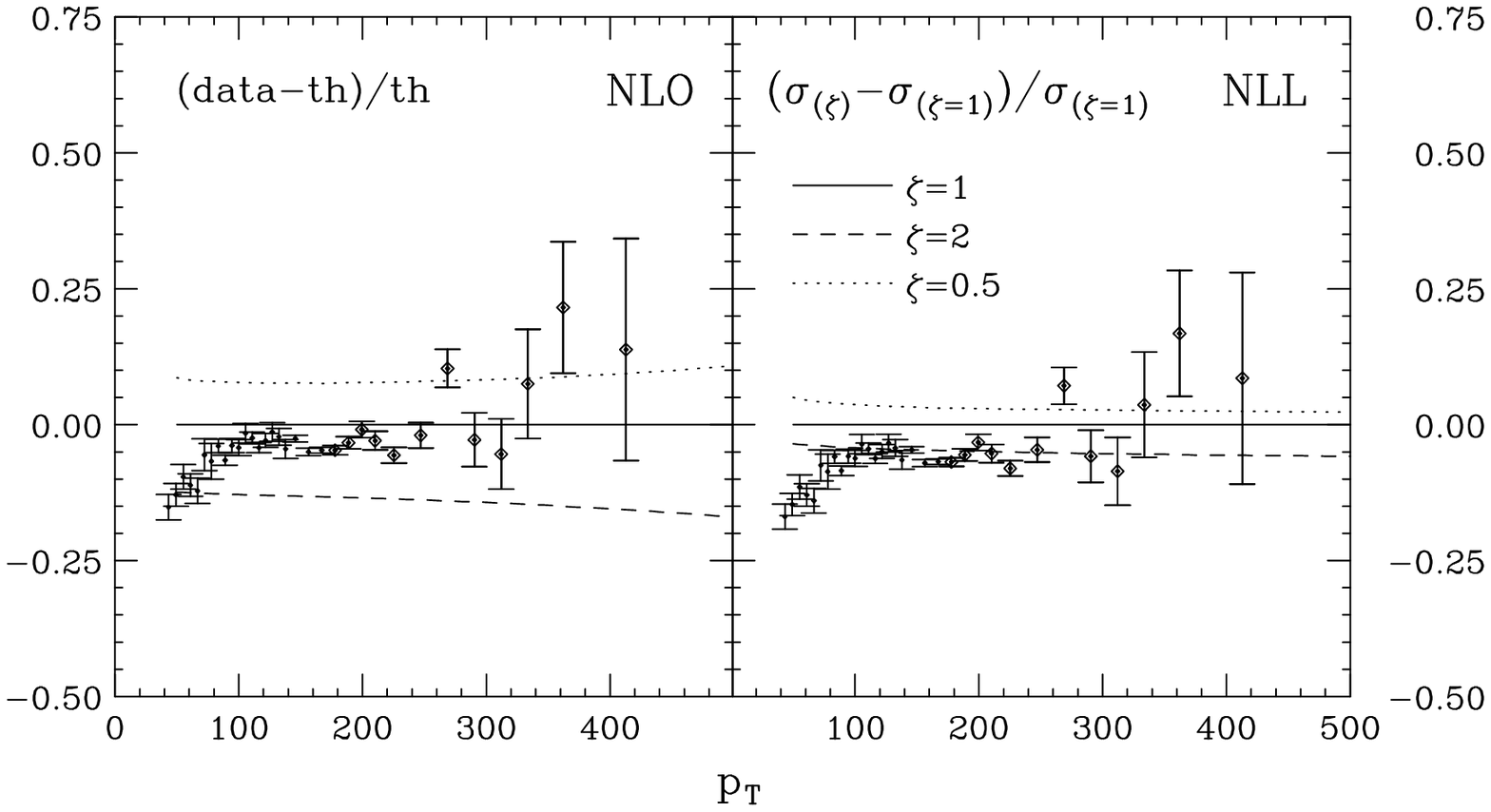,width=0.8\textwidth}
\end{center}
\vspace*{-.5cm}
\caption{Ratios ``(data$-$theory)/theory'' for data from 
CDF at 1.8~TeV~\cite{Affolder:2001fa} and for NLO and NLL resummed
theoretical calculations. We have chosen the theory result at 
scale $p_T$ as ``default''; results for other scales are also 
displayed in terms of their relative shifts with respect to the
default theory.    \label{fig:a5}
\label{dminust}}
\vspace*{.6cm}
\end{figure}

Finally, as one illustrative example, we compare our resummed jet 
cross section to data from CDF~\cite{Affolder:2001fa} at 
$\sqrt{S}=1800$~GeV. While so far
we have always considered the cross section integrated over
all jet rapidities, we here need to account for the fact that the 
data cover only a finite region in rapidity, $0.1 \leq |\eta|\leq 0.7$. Also, 
we would like to properly match the jet algorithm chosen in experiment, 
rather than using the SCA. We have mentioned before that both
these issues can be accurately addressed by ``rescaling''
the resummed cross section by an appropriate ratio of 
NLO cross sections. We simply multiply our $K$-factors defined in Eq.~(\ref{eq:kres}) and shown 
in Fig.~\ref{fig:a3} by  $d\sigma^{\rm (MC)}(0.1 \leq |\eta|\leq 0.7)
/dp_T$, 
 the NLO cross section obtained with a full Monte-Carlo 
code~\cite{Frixione:1997np}, in the experimentally accessed
rapidity regime.
The comparison between the data and the NLO and NLL-resummed cross
sections is shown in Fig.~\ref{dminust} in terms of the ratios
``(data$-$theory)/theory''. As expected from Fig.~\ref{fig:a3},
the impact of resummation is moderate and in fact smaller than the current
uncertainties of the parton distributions~\cite{tevrun2}. Nonetheless, 
it does lead to a slight improvement of the comparison and,
in particular, the plot again demonstrates the reduction of scale
dependence by resummation.

\section{Conclusions \label{sec5}}

We have studied in this paper the resummation of large logarithmic 
threshold corrections to the partonic cross sections contributing to
single-inclusive jet production at hadron colliders. Our study differs 
from previous work~\cite{KO} mostly in that we allow the jet to have
a finite invariant mass at partonic threshold, which is consistent 
with the experimental definitions of jet cross sections and with the 
available NLO calculations. Moreover, using semi-analytical expressions
for the NLO partonic cross sections derived in the 
SCA~\cite{Aversa:1988vb,Jager:2002xm}, we have extracted 
the $N-$independent coefficients that appear in the resummation formula,
and properly matched our resummed cross section to the NLO one. We hope
that with these improvements yet more realistic estimates of the higher-order
corrections to jet cross sections in the threshold regime emerge.

It is well known that the NLO description of jet production at
hadron colliders is overall very successful, within the uncertainties
of the theoretical framework and the experimental data. From that
perspective, it is gratifying to see that the effects of NLL resummation
are relatively moderate. On the other hand, resummation leads to a
significant decrease of the scale dependence, and we expect that
knowledge of the resummation effects should be useful in comparisons 
with future, yet more precise, data, and for extracting parton distribution 
functions. Given the general success of the NLO description, we have 
mostly focused on $K$-factors for the resummed cross section over the NLO 
one, and only given one example of a more detailed comparison with 
experimental data.
We believe that these $K$-factors may be readily used in conjunction
with other, more flexible NLO Monte-Carlo programs for jet production,
to estimate threshold-resummation effects on cross sections for other
jet algorithms and possibly for larger cone sizes.

\section*{Acknowledgments}
We are grateful to Stefano Catani, Barbara J\"{a}ger, Nikolaos Kidonakis, 
Douglas Ross, George Sterman, and
Marco Stratmann for helpful discussions.
DdF is supported in part by UBACYT and CONICET.
WV~is supported by the U.S. Department of Energy under contract number 
DE-AC02-98CH10886.

\section*{Appendix: First-order coefficients $C_{ab}^{(1)}$ in the SCA}

In this appendix we collect the process-dependent coefficients 
$C_{ab}^{(1)}$ for the various partonic channels in jet hadroproduction 
in the SCA. The $C_{ab}^{(1)}$ are constant, that is, they 
do not depend on the 
Mellin-moment variable $N$. They may be extracted by expanding the resummed 
cross section in Eq.~(\ref{eq:res}) to first order and comparing it to 
the full NLO cross section in the SCA. For the sake of simplicity, we 
provide the $C_{ab}^{(1)}$ only in numerical form, as the full analytic 
coefficients have relatively lengthy expressions. We find:
\begin{eqnarray}
C^{(1)}_{qq'}&=& 17.9012 + 1.38763 \log\frac{R}{2} \, , \nonumber \\
C^{(1)}_{q\bar{q'}}&=& 19.0395 + 1.38763 \log\frac{R}{2} \, , 
\nonumber \\
C^{(1)}_{q\bar{q}}&=& 13.4171 + 1.6989 \log\frac{R}{2}\, , 
\nonumber \\
C^{(1)}_{qq}&=&17.1973 + 1.38763 \log\frac{R}{2} \, , 
\nonumber \\
C^{(1)}_{qg}&=& 14.4483 + 2.58824 \log\frac{R}{2} \, , 
\nonumber \\
C^{(1)}_{gg}&=& 14.5629 + 3.67884 \log\frac{R}{2} \, ,
\end{eqnarray}
where $R$ is the jet cone size.

\end{document}